\documentclass{article}

\newcommand{\Tr}{{\rm Tr}}

\newcommand{\be}{\begin{equation}}
\newcommand{\ee}{\end{equation}}

\begin{document}

\title{Time delay statistics for chaotic cavities with absorption}
\author{Marcel Novaes\\Instituto de F\'isica, Universidade Federal de Uberl\^andia \\Uberl\^andia, MG, 38408-100, Brazil}

\maketitle

\abstract{We present a semiclassical approach for time delay statistics in quantum chaotic systems, in the presence of absorption, for broken time-reversal symmetry. We derive three kinds of expressions for Schur-moments of the time delay operator: as a power series in inverse channel number, $1/M$, whose coefficients are rational functions of absorption time, $\tau_a$; as a power series in $\tau_a$, tailored to strong absorption, whose coefficients are rational functions of $M$; as a power series in $1/\tau_a$, tailored to weak absorption, whose coefficients are rational functions of $M$.}

\section{Introduction}

We consider quantum scattering in a cavity, connected to the outside by a finite number of scattering channels, $M$, determined by transverse quantisation in a wave guide connected to a cavity. Amplitudes of outgoing waves are then related to those of incoming waves through multiplication by an $M$-dimensional $S$ matrix, which must be unitary when there is no absorption in order to guarantee energy conservation. This setup is quite general and finds numerous applications \cite{a1,a2,a3,a4,a5,a6}. We shall assume the classical dynamics to be chaotic at all energies and characterized by a well defined dwell time $\tau_d$, the average time spent in the scattering region. We also assume broken time-reversal symmetry.

When the scattering matrix is unitary, its energy derivative defines what is known as the Wigner–Smith time delay matrix \cite{time1,time2,time3,time4,time5},
\be\label{Qd} Q=-i\hbar S^\dagger \frac{dS}{dE},\ee
where $\hbar$ is Planck's constant. This is an operator representing the quantization of the notion of time delay, i.e. the time spent by quantum particles inside the scattering region. As such, it is of importance in different areas \cite{t0,t1,t2,t3,t4,t5,t6}. The normalized trace $\tau_W = \frac{1}{M}\Tr(Q)$ is called the Wigner time delay. Its related to the density of states of the open system \cite{d,rmt0} and its average value is the classical dwell time, $\langle\tau_W \rangle=\tau_d$. More refined statistical information about the time duration of wave scattering is encoded in other spectral properties of $Q$.

For complex or chaotic systems, the matrix elements of $Q$ are typically widely fluctating functions of the energy. Within a random matrix theory (RMT) approach, detailed characteristics of the system are ignored and instead $Q$ is considered as random \cite{rmt0,rmt1,rmt2,rmt3,rmt4}. This is a traditional and very fruitful point of view, that has lead to a great number of results \cite{results2,results3,results4,cunden,results5,results6,results7,results8,
eu1,grabsch,eu2}. 

Another possibility is to employ a semiclassical approximation in which the elements of $S$ are written as infinite sums over scattering rays. Then, for completely chaotic systems, local energy averages can be introduced which lead to some integrations over phase space, out of which a set of diagrammatic rules come out \cite{semi1,semi2,semi3,delay0,delay1,delay2,delay3,efficient}, which can be used to derive statistical properties of time delay.

Both these approaches are very succesfull when absorption is disregarded. However, the phenomenom of absorption is always present to some level and must be taken into account for a proper description of some experiments. The strength of the absorption is characterised by the absorption time $\tau_a$, the mean
time a wave can spend in the system before being absorbed. The ratio $\tau_d/\tau_a$ then becomes the relevant parameter. In such a context, the scattering matrix is no longer unitary and the definition (\ref{Qd}) leads to a non-hermitian operator with complex eigenvalues. Real and imaginary parts of such complex time delay have in fact recently attracted much attention \cite{c1,c2,c3,c4}. 

On the other hand, a different notion of time delay has been defined for systems with absorption by measuring the unitarity deficit of the scattering matrix \cite{abs1}, according to 
\be Q(\gamma)=\frac{1}{\gamma}(1_M-S S^\dagger),\ee 
where $1_M$ is the $M$-dimensional identity matrix and 
\be \gamma=\frac{\tau_d}{\tau_a}.\ee This formulation has lead to progress on the RMT side \cite{abs2,abs3,abs4}. Savin and Sommers found that 
\be \langle \tau_W\rangle =\frac{\tau_d}{1+\gamma}+O(M^{-1}).\ee Grabsch found the joint distribution of eigenvalues of $Q$, valid for any $M$, and used a Coulomb gas technique to show that, when $M\gg 1$,
\be {\rm var}(\tau_W)\approx\frac{2(1-6\gamma)}{M^2}\tau_d^2,\ee
in the regime of weak absorption, $\gamma\ll 1$, and
\be{\rm var}(\tau_W)\approx\frac{1}{ M^2\gamma^4}\tau_d^2,\ee
in the regime of strong absorption, $\gamma\gg 1$.

In this work we treat absorption within a semiclassical approach, by taking advantage of diagrammatic rules obtained in \cite{delay2}, further discussed in \cite{delay3}, implemented systematically by means of the matrix integral method developed in \cite{matrix1,matrix2,matrix3}. This method is able to go beyond RMT, for example when it is necessary to take into account energy dependence \cite{matrix4,matrix5,matrix6} or the presence of tunnel barriers \cite{tunnel1,tunnel2,tunnel3,tunnel4}. Moreover, it allows for results that are exact in $M$. For example, we obtain that
\be \frac{\langle \tau_W\rangle}{\tau_d} =\frac{1}{1+\gamma}-\frac{\gamma}{M^2(1+\gamma)^5}-\frac{\gamma(8\gamma^2-12\gamma+1)}{M^4(1+\gamma)^9}+O(M^{-6}),\ee
\be \frac{\langle \tau_W\rangle}{\tau_d} =1+\frac{\gamma M^2}{M^2-1}+\frac{\gamma^2 M^4}{(M^2-1)(M^2-4)}+O(\gamma^3),\ee
and
\be \frac{\langle \tau_W\rangle}{\tau_d} =\frac{1}{\gamma}-\frac{1}{\gamma^2}+\frac{1}{\gamma^3}-\frac{M^2+1}{M^2\gamma^4}+\frac{M^2+5}{M^2\gamma^5}+O(\gamma^{-6}),\ee
for the average Wigner time delay. For its variance, we obtain
\be \frac{{\rm var}(\tau_W)}{\tau_d^2}= \frac{\gamma^2+2}{M^2(1+\gamma)^6}+\frac{2-40\gamma+68\gamma^2-28\gamma^3+8\gamma^4}{M^4(1+\gamma)^{10}}+O(M^{-6}),\ee
\be \frac{{\rm var}(\tau_W)}{\tau_d^2}= \frac{2}{M^2-1}-\frac{12\gamma M^2}{(M^2-1)(M^2-4)}+O(\gamma^2),\ee
\be \frac{{\rm var}(\tau_W)}{\tau_d^2}= \frac{1}{M^2\gamma^4}-\frac{6}{M^2\gamma^5}+\frac{23M^2+8}{M^4\gamma^6}+O(\gamma^{-7}).\ee

Other results and a few conjectures are presented in the following sections. We start by discussing the semiclassical approximation in Section 2, and then we derive three different kinds of explicit results for time delay statistics: as a power series in inverse channel number, $1/M$, whose coefficients are rational functions of $\gamma$; as a power series in $\gamma$, appropriate for when absorption is weak, whose coefficients are rational functions of $M$; as a power series in $\gamma^{-1}$, appropriate for when absorption is strong, whose coefficients are rational functions of $M$. We conclude in Section 6.

\section{Semiclassical approach}

The effects of absorption can be incorporated into the theory by adding an imaginary part to the energy, which is inversely proportional to the absorption time, i.e. taking it in the form
\be E+\frac{i\hbar}{2\tau_a}=E+\frac{i\hbar\gamma}{2\tau_d}=E+\epsilon'.\ee
Then the scattering matrix is no longer unitary, and 
\be R=S(E+\epsilon') S^\dagger(E-\epsilon')\ee is a reflection matrix. The spectral properties of $Q(\gamma)=(1_M-R)/\gamma$ are of course related to the spectral properties of $R$. 

Given an integer partition $\lambda=(\lambda_1,\ldots,\lambda_{\ell(\lambda)})$, a non-decreasing sequence of $\ell(\lambda)$ positive integers such that 
\be \sum_i \lambda_i=|\lambda|,\ee let $s_\lambda(Q)$ be the corresponding Schur polynomial. This is a family of symmetric polynomials in the eigenvalues of $Q$ which form a basis for the vector space of homogeneous symmetric polynomials of degree $|\lambda|$. For example, we can write $\tau_W=s_{(1)}(Q)$ and $\tau_W^2=\frac{1}{2}s_{(2)}(Q)+\frac{1}{2}s_{(1,1)}(Q)$. 

More familiar quantities like ${\rm Tr}(Q^n)$ or $[{\rm Tr}(Q)]^n$ may be expressed in the form of power sum symmetric polynomials,
\be p_\lambda(Q)=\prod_{i=1}^{\ell(\lambda)}{\rm Tr}(Q^{\lambda_i}),\ee which in turn are written as linear combinations of Schur polynomials as \cite{stan}
\be p_\lambda(Q)=\sum_\mu \chi_\mu(\lambda)s_\mu(Q),\ee where the coefficient $\chi_\mu(\lambda)$ is a character of the symmetric group in the irreducible representation labelled by $\mu$ computed at a permutation of cycle type $\lambda$. 

We call the local energy averages $\langle s_\lambda(Q)\rangle$ the Schur-moments of $Q$. They can be written in terms of Schur polynomials in $R$ according to the generalized binomial theorem,
\be\label{bin} s_\lambda(Q)=\frac{1}{\gamma^{|\lambda|}}\sum_{\mu\subset\lambda} (-1)^{|\mu|}B_{\lambda\mu}s_\mu(R),\ee where the coefficients are given by a determinant involving binomials,
\be B_{\lambda\mu}=\det \left[ {M+\lambda_i-i \choose M+\mu_j-j}\right].\ee

The semiclassical approach proceeds by writing the matrix elements of $R$ in terms of pairs of scattering trajectories. Spectral statistics then involve multiple sums over these trajectories. Under a local energy average, a traditional stationary phase approximation can be invoked, according to which constructive interference requires action correlations. Summation over trajectories becomes integration over phase space, which lead to a diagrammatic formulation: calculation of moments can be done in terms of a sum over ribbon graphs, with weights attached to vertices and edges. This approach has been used to obtain transport properties \cite{semi1,semi2,semi3} and time delay statistics \cite{delay2,delay3,efficient}, in a variety of regimes. 

Calculation of Schur moments $\langle s_\mu(R)\rangle$, in particular, was developed in detail in \cite{matrix6}, and we shall rely on those results. That work assumed $\epsilon'$ to be real, but this restriction can be lifted without any difficulty. 

Let us define a generalization of the rising factorial,
\be [M]^\mu=\prod_{i=1}^{\ell(\mu)}\frac{(M+\mu_i-i)!}{(M-i)!}.\ee
The method developed in \cite{matrix6} then leads to the result
\be\label{s} \langle s_\mu(R)\rangle=\lim_{N\to 0}\frac{([M]^\mu)^2}{([N]^\mu)^2}I_\mu,\ee in terms of an integral over $N$-dimensional positive real diagonal matrices $X$,  
\be\label{int} I_\mu(M,\gamma,N)=\int e^{-M\sum_{q=1}^\infty (1/q+\gamma){\rm Tr}(X^q)} |\Delta(X)|^2s_\mu(X)dX,\ee
where $\Delta(X)=\prod_{k>j}(x_k-x_j)$ is the Vandermonde of the eigenvalues of $X$. The diagrammatical formulation of this integral is equivalent to the semiclassical diagrammatics. But notice how the dimension of $X$ is taken to $0$ after the integral has been computed. This somewhat unusual procedure is necessary in order to avoid the presence of spurious periodic orbits in the semiclassical diagrams \cite{matrix1,matrix5}. 

When taking the limit $N\to 0$, we shall use that $[N]^\mu\approx t_\mu N^{D(\mu)}$ when $N$ is small. The quantities $t_\mu$ and $D(\mu)$ that appear in this expression are simply derived from the partition $\mu$: $D(\mu)$ is the side of the Durfee square of $\mu$, the largest square that fits in the Young diagram of $\mu$, while $t_\mu$ is the product of all non-zero contents of $\mu$ (see \cite{matrix4}). 

The integral in (\ref{int}), together with (\ref{s}) and (\ref{bin}), provides in principle an exact representation for the Schur moments of time delay, valid for any value of the parameters $M,\gamma$. However, computing it explicitly is a challenge, and we cannot do it in closed form. Some approximation is necessary. We develop three different asymptotic expansions in the next sections. One for $M\gg 1$, one for $\gamma\ll 1$ and one for $\gamma\gg 1$. The first one treats $\gamma$ exactly, while the other two treat $M$ exactly.

In the following we shall use for simplicity $\tau_d=1$, i.e. we will measure all times in units of the classical dwell time (this is in contrast with \cite{abs2}, which uses the Heisenberg time $M\tau_d$ as unit).

\section{Expression in powers of $1/M$}

We can leave the term $q=1$ in exponential form, and expand the rest of the exponential in powers of $M$. This calculation can be found in \cite{matrix6}, so we present only a sketch of the argument and the final result. 

Let $b_\beta$ be the size of the conjugacy class of the permutation group containing permutations of cycle type $\beta$ (the largest cycle has length $\beta_1$, the second largest has length $\beta_2$ etc) and define 
\be g_\beta(\gamma)=\prod_{q\in\beta}(1+q\gamma).\ee
Let $c^\nu_{\mu\rho}$ be the Littlewood-Richardson coefficients in the expansion 
\be s_\rho(X)s_\mu(X)=\sum_\nu c^\nu_{\mu\rho}s_\nu(X).\ee Then, we have
\be I_\mu =\sum_{m} \sum_{\beta\vdash m}{}^{'}\frac{b_\beta}{m!}(-M)^{\ell(\beta)}g_\beta(\gamma)\sum_{\rho\vdash m} \chi_\rho(\beta)\sum_{\nu\vdash |\mu|+m} c^\nu_{\mu\rho} j_\nu,\ee
where the partitions $\beta$ being summed over have no parts equal to $1$. The integral to be done is
\be\label{Ie} j_\nu=\int e^{-M(1+\gamma)\Tr(X)}|\Delta(X)|^2s_\nu(X)dX,\ee
which is of Selberg type \cite{selberg}. Defining $d_\nu=\chi_\nu(1)$ to be the dimension of irreducible representations of the symmetric group, this integral is given by
\be \frac{d_\nu}{|\nu|!}([N]^\nu)^2[M(1+\gamma)]^{N^2-|\nu|}.\ee 

After the limit $N\to 0$ is taken \cite{matrix1,matrix5}, we have 
\be \langle s_\mu(R)\rangle=\frac{([M]^\mu)^2}{t_\mu^2}\sum_{m} \sum_{\beta\vdash m}{}^{'}\frac{F_\mu^{(1)}(\gamma,\beta)}{M^{|\mu|+m-\ell(\beta)}},\ee
where 
\be F_\mu^{(1)}(\gamma,\beta)=\frac{(-1)^{\ell(\beta)}b_\beta g_\beta(\gamma)}{m!(n+m)!(1+\gamma)^{n+m}}\sum_\nu c^\nu_{\mu\rho} d_\nu t_\nu^2\delta_{D(\nu),D(\mu)}.\ee The Kronecker delta above requires $\nu$ and $\mu$ to have the same Durfee square.

The above expression, which easily leads to results as power series in $M^{-1}$, is suited to the regime of many open channels, when $M\gg 1$. 

For example, we have that $\frac{1}{M}\langle{\rm Tr}(Q^2)\rangle$ equals
\be \frac{\gamma^2+2\gamma+2}{(1+\gamma)^4}-\frac{4\gamma^3-\gamma^2+14\gamma-2)}{(1+\gamma)^8M^2}+O(M^{-4})\ee
and that $\frac{1}{M^2}\langle[{\rm Tr}(Q)]^{2}\rangle$ equals
\be \frac{1}{(1+\gamma)^2}+\frac{\gamma^2-2\gamma+2}{(1+\gamma)^6M^2}+\frac{8\gamma^4-44\gamma^3+93\gamma^2-42\gamma+2}{(1+\gamma)^{10}M^4}+O(M^{-6})\ee

General expressions are quite cumbersome, but experimentation with computer algebra suggests that $\frac{1}{M^n}\langle[{\rm Tr}(Q)]^{n}\rangle$ equals
\be \frac{1}{(1+\gamma)^n}+\frac{n(n-1)\gamma^2/2-n\gamma+n(n-1)}{(1+\gamma)^{n+4}M^2}+O(M^{-4}),\ee 
for any positive integer $n$.

Concerning cumulants of the Wigner time delay, we have shown the first two cumulants in the Introduction, the next one, $k_3$, is given, to leading order in $1/M$, by
\be -\frac{2\gamma^8-8\gamma^7+26\gamma^6-48\gamma^5-348\gamma^4-420\gamma^3-221\gamma^2-64\gamma-8}{\gamma^3(1+\gamma)^{11}M^4}.\ee

\section{Expression in powers of $\gamma$}

If we sum both series in the exponent, we get
\be\label{Ig} I_\mu(N,\gamma)=\int \det(1-X)^Me^{-M\gamma{\rm Tr}\left(\frac{X}{1-X}\right)}|\Delta(X)|^2 s_\mu(X)dX.\ee The exponential can now be expanded in powers of $\gamma$, with coefficients that involve $s_\rho\left(\frac{X}{1-X}\right)$. Then the integral is computed using the Andreief identity. This calculation can be found in \cite{matrix6}, so we present only the final result, which is
\be\label{f2} \langle s_\mu(R)\rangle=\frac{[M]^\mu}{t_\mu^2}\sum_{m=0}^\infty F_\mu^{(2)}(M,m)\gamma^m,\ee where
\be F_\mu^{(2)}(M,m)=\frac{(-M)^m}{m!(n+m)!}\sum_{\rho\vdash m} \frac{d_\rho}{[M]_\rho}\sum_\nu c^\nu_{\mu\rho} d_\nu t_\nu^2\delta_{D(\nu),D(\mu)}\ee
involves a generalization of the falling factorial,
\be [M]_\rho=\prod_{i=1}^{\ell(\rho)}\frac{(M+i-1)!}{(M+i-\rho_i-1)!}.\ee

The expression (\ref{f2}), written as a power series in $\gamma$, is suited to the regime of weak absorption, when $\tau_a\gg \tau_d$. 

For example, we have that $\frac{1}{M}\langle{\rm Tr}(Q^2)\rangle$ equals
\be \frac{2M^2}{M^2-1}-\frac{6\gamma M^4}{(M^2-1)(M^2-4)}+O(\gamma^2)\ee
and that $\frac{1}{M^2}\langle[{\rm Tr}(Q)]^{2}\rangle$ equals
\be \frac{M^2+1}{M^2-1}-\frac{2\gamma M^2(M^2+2)}{(M^2-1)(M^2-4)}+O(\gamma^2).\ee

General expressions are again quite cumbersome, but experimentation with computer algebra suggests an interesting relation. Let $P_n(\gamma)=\frac{1}{M}\langle{\rm Tr}(Q^n)\rangle$. Then,
\be P_n(\gamma)=P_n(0)-\frac{n}{2}P_{n+1}(0)\gamma+O(\gamma^2).\ee 

A somewhat similar relation also seems to hold for cumulants of the Wigner time delay,
\be\label{k} k_n(\gamma)=k_n(0)-\frac{M^2}{2}k_{n+1}(0)\gamma+O(\gamma^2).\ee 
We conjecture that these relations hold for finite $M$ (relation (\ref{k}) was proved by Grabsch \cite{abs2} in the regime $M\gg 1$). For example,
\be k_3=\frac{24}{(M^2-1)(M^2-4)}-\frac{6M^2(53M^2-77)\gamma}{(M^2-1)^2(M^2-4)(M^2-9)}+O(\gamma^2)\ee
and
\be k_4=\frac{12(53M^2-77)\gamma}{(M^2-1)^2(M^2-4)(M^2-9)}+O(\gamma).\ee

\section{Expression in powers of $1/\gamma$}

This calculation was not done is \cite{matrix6}, so we provide a bit more detail. We start by changing variables in (\ref{Ig}) as
\be X=\frac{Y}{1+Y}.\ee
This leads to 
\be \Delta(X)=\frac{\Delta(Y)}{\det(1+Y)^{N-1}}\ee
and $dX=\det(1+Y)^{-2}dY.$ Hence,
\be I_\mu(N,\gamma)=\int \frac{e^{-M\gamma{\rm Tr}(Y)}}{\det(1+Y)^{M+2N}}|\Delta(Y)|^2 s_\mu\left(\frac{Y}{1+Y}\right)dY.\ee 

The determinant is expanded using the Cauchy identity,
\be \frac{1}{\det(1+Y)^{M+2N}}=\sum_\omega (-1)^{|\omega|}s_\omega(1_{M+2N})s_\omega(Y).\ee This is an infinite sum over partitions which generalized the geometric series. On the other hand, the Schur polynomial is expanded as another generalization of the geometric series,
\be s_\mu\left(\frac{Y}{1+Y}\right)=\sum_{\rho \supset\mu}(-1)^{|\rho|+|\mu|}G_{\mu\rho}s_\rho(Y),\ee
where
\be G_{\mu\rho}=\det \left[ {\rho_i-i \choose \mu_j-j}\right].\ee

Using the Littlewood-Richardson coefficients, we again arrive an integral of the kind (\ref{Ie}), and finally at 
\be \langle s_\mu(R)\rangle=\frac{(-1)^{|\mu|}([M]^\mu)^2}{t_\mu^2}\sum_{\rho \supset\mu}\sum_\omega \frac{F_\mu^{(3)}(M,\rho,\omega)}{\gamma^{|\omega|+|\rho|}},\ee
where 
\be F_\mu^{(3)}(M,\rho,\omega)=\frac{d_\omega [M]^\omega G_{\mu\rho}}{|\omega|!(-M)^{|\omega|+|\rho|}}\sum_\nu c^\nu_{\omega\rho} d_\nu t_\nu^2\delta_{D(\nu),D(\mu)}.\ee

The above expression, written as a power series in $\gamma^{-1}$, is suited to the regime of strong absorption, when $\tau_a\ll \tau_d$. 

For example, we have that $\frac{1}{M}\langle{\rm Tr}(Q^2)\rangle$ equals
\be \frac{1}{\gamma^2}-\frac{2}{\gamma^3}+\frac{4}{\gamma^4}-\frac{4(2M^2+1)}{\gamma^5M^2}+O(\gamma^{-6})\ee
and that $\frac{1}{M^2}\langle[{\rm Tr}(Q)]^{2}\rangle$ equals
\be \frac{1}{\gamma^2}-\frac{2}{\gamma^3}+\frac{3M^2+1}{\gamma^4 M^2}-\frac{4(M^2+2)}{\gamma^5M^2}+O(\gamma^{-6}).\ee
We conjecture that $\frac{1}{M}\langle{\rm Tr}(Q^n)\rangle$ is given by
\be \frac{1}{\gamma^n}-\frac{n}{\gamma^{n+1}}+\frac{n^2}{\gamma^{n+2}}-\frac{(n+1)[n(5n-2)M^2+n(n+2)]}{6M^2\gamma^{n+3}}+O(\gamma^{-n-4})\ee
and that $\frac{1}{M^n}\langle[{\rm Tr}(Q)]^{n}\rangle$ is given by
\be \frac{1}{\gamma^n}-\frac{n}{\gamma^{n+1}}+\frac{n[(n+1)M^2+(n-1)]}{2M^2\gamma^{n+2}}+O(\gamma^{-n-3}),\ee
for every positive integer $n$.

As for the cumulants of the Wigner time delay, we have
\be k_3=-\frac{2}{M^4\gamma^6}+\frac{30}{M^4\gamma^7}-\frac{6(41M^2+10)}{M^6\gamma^8}+O(\gamma^{-9})\ee
and
\be k_4=\frac{6}{M^6\gamma^8}-\frac{168}{M^6\gamma^9}+O(\gamma^{-10})\ee
we conjecture that, for $n>1$,
\be (-1)^nk_n(\gamma)=\frac{(n-1)!}{M^{2n-2}\gamma^{2n}}-\frac{(2n-1)n(n-1)!}{M^{2n-2}\gamma^{2n+1}}+O(\gamma^{-2n-2}).\ee 

\section{Conclusion}

We have used a semiclassical approximation, combined with the machinery of matrix integrals, to obtain several time delay statistics for scattering through quantum chaotic systems, in the presence of absorption, for broken time-reversal symmetry. 

Our results consist in three kinds of expressions for Schur-moments of the time delay operator. When there are many open channels, we can write them as power series in $1/M$, whose coefficients are rational functions of the absorption strength, $\gamma$. For weak absorption, we can write them as power series in $\gamma$ with coefficients that are rational functions of $M$. Finally, for strong absorption we can write them as power series in $1/\gamma$, again with coefficients that are rational functions of $M$. These results go beyond what was previously available from random matrix theory approaches.

The calculations we have carried out can in principle also be used for systems with intact time-reversal symmetry, by using zonal polynomials is place of Schur polynomials. We leave that topic for future work.

Another important generalization would be to take into account the possibility of imperfect coupling between the system and the outside, for instance in the form of tunnel barriers. Some results combining absorption with imperfect coupling are already available, both from random matrix theory \cite{abs1,abs4} and semiclassical theory \cite{kui1,kui2}, but so far only the regime $M\gg 1$ has been addressed. 

Data sharing not applicable to this article as no datasets were generated or analysed during the current study.

The author has no competing interests to declare that are relevant to the content of this article.

This research had financial support from Fapemig, under grant PPM-00126-17.

\end{document}